\documentclass[a4paper,fleqn,usenatbib]{mnras}

\usepackage{newtxtext,newtxmath}

\usepackage[T1]{fontenc}
\usepackage{ae,aecompl}

\usepackage{graphicx}	
\usepackage{amsmath}	
\usepackage{amssymb}	

\title[FIR metallicity diagnostics for high-z galaxies]{On the far-infrared
  metallicity diagnostics: applications to high-redshift galaxies   }

\author[D. Rigopoulou et al.]{
D. Rigopoulou,$^{1}$\thanks{E-mail:dimitra.rigopoulou@physics.ox.ac.uk}
M. Pereira-Santaella$^{1}$,
G.E. Magdis$^{2}$,
A. Cooray$^{3}$,
D. Farrah$^{4}$,
\newauthor
R. Marques-Chaves$^{5,6}$,
I. Perez-Fournon$^{5,6}$,
D. Riechers$^{7}$
\\
$^{1}$Astrophysics, Department of Physics, University of Oxford, Keble
Road, Oxford, OX1 3RH, UK\\
$^{2}$Dark Cosmology Centre, Niels Bohr Institute, University of
Copenhagen, Juliane Mariesvej 30, 
DK-2100 Copenhagen, Denmark\\
$^{3}$Department of Physics \& Astronomy, University of California,
Irvine, CA 92697, USA\\
$^{4}$Department of Physics, Virginia Tech, Blacksburg, VA 24061,
USA\\
$^{5}$Instituto de Astrofisica de Canarias, E-38205 La Laguna, Tenerife, Spain\\
$^{6}$Universidad de La Laguna, Dpto. Astrofisica, E-38206 La Laguna, Tenerife, Spain\\
$^{7}$Department of Astronomy, Cornell University, 220 Space Sciences Building, Ithaca, NY 14853, USA\\
}
\date{Accepted XXX. Received YYY; in original form ZZZ}

\pubyear{2017}

\begin{document}
\label{firstpage}
\pagerange{\pageref{firstpage}--\pageref{lastpage}}
\maketitle

\begin{abstract}
In an earlier paper we modeled the far-infrared emission from a star-forming galaxy using the photoionization code
CLOUDY and  presented metallicity sensitive diagnostics based on
far-infrared fine structure line ratios. Here, we focus on the applicability
of the [OIII]88 {$\rm\mu$}m$/$[NII]122 {\rm$\mu$}m
line ratio as a gas phase metallicity indicator in high redshift submillimetre luminous galaxies.
The [OIII]88 $\mu$m$/$[NII] 122$\mu$m ratio is strongly dependent on the ionization
parameter (which is related to the total
number of ionizing photons) as well as the gas electron density. We demonstrate
how the ratio of 88 $\mu$m$/$122 $\mu$m continuum flux measurements can provide a
reasonable estimate of the ionization parameter while the availability
of the [NII]205 $\mu$m line can
constrain the electron density.
Using the [OIII]88 $\mu$m$/$[NII]122 $\mu$m line ratios from a sample
 of nearby normal and star-forming galaxies we measure their gas phase metallicities
and find that their mass metallicity relation is consistent with the one derived
using optical emission lines. Using new, previously unpublished, Herschel spectroscopic 
observations of key far-infrared fine structure
lines of the z$\sim$3 galaxy HLSW-01 and additional published
measurements of  far-infrared fine
structure lines of high-z submillimetre luminous galaxies 
we derive gas phase metallicities using  their 
[OIII]88 $\mu$m$/$[NII]122 $\mu$m
line ratio. We find that the metallicites of
these  z$\sim$3 submm luminous galaxies are
consistent with solar metallicities and that  they
appear to follow the mass-metallicity relation expected  for z$\sim$3 systems.

\end{abstract}




\section{Introduction}

The metallicity of a galaxy is closely linked to its star formation
history and the inflow and outflow of gas, the interplay between the
galaxy's insterstellar medium (ISM) and the intergalactic medium
(IGM). Gas accretion from the IGM to the ISM together with large
scale outflows are the two principal components in galaxy formation 
and evolution models (e.g. Dekel et al. 2009, Dav\'e et al. 2012). 
Feedback from star formation drives outflows that remove mass and
metals away from galaxies, whereas, infall of gas from the IGM to the
ISM is necessary to sustain star-formation. Without the feedback,
baryons would cool into the centers of haloes and form prodigious
amounts of stars (e.g. Keres et al. 2009) but with feedback the
baryonic content of stars and cold gas in galaxies can be reconciled by
driving matter into the IGM (e.g. Conroy \& Wechsler 2009). Likewise,
without infall of material from the  IGM star forming galaxies would
use up their ISM gas in $\sim$1 Gyr (e.g. Genzel et al. 2010).
To understand these processes we need to study how
galaxies obtain, process, expel and recycle gas from their
surroundings.

Galaxies evolve through the build-up of stellar mass (M$_{*}$)
that takes place over time through a number of star formation
episodes. As stars evolve they produce metals so that the metal
content of a galaxy, typically measured by the oxygen abundance (O$/$H),
 can be used as a diagnostic for tracing galaxy formation
and evolution. The presence of a tight correlation between the stellar mass of a
galaxy and its gas-phase metallicity
(Z), the so called M$_{*}$-Z relation
has been found to hold in the nearby Universe (Tremonti et
al. 2004) and up to redshifts z$\sim$2 (e.g. Erb et al. 2006,
Maiolino et al. 2008, Wuyts et al. 2014) and z$\sim$3 (e.g. Mannucci
et al 2009, Onodera et al. 2016). Mannucci et al. (2010) extended this
correlation to include the star formation as well, defining the
so-called fundamental metallicity relation (FMR).

Gas phase galaxy  metallicities have traditionally been derived using
 ratios between strong optical  emission lines
 ([OII]$\lambda$3727\.A, [OIII]$\lambda$5007\.A, H$_{\alpha}$,
 H$_{\beta}$, 
[NII]$\lambda\lambda$6549\.A,6583\.A)
calibrated on theoretical models (e.g., Kewley \& Dopita 2002;
Kewley \& Ellison 2008; Maiolino et al. 2008).
 The most commonly used ones are those based on the  ``strong line
methods'' and include the R23 diagnostics (combining [OIII],
[OII], and H$_{\beta}$; Pagel et al. 1979) and the N2 method
(combining H$_{\alpha}$ and [N II]; Storchi-Bergmann et al. 1994). The
N2 method is commonly used to break the degeneracy and dust
dependency of the R23 method. The strong-line methods
can be used for galaxies up to z $\sim$3, at higher redshifts however, the diagnostic
lines shift out of the wavelengths accessible by ground-based near-
IR spectrographs. A correlation between metallicity and the equivalent
width (EW) of absorption features in the rest-frame ultraviolet (UV)
is expected based on theoretical models (e.g. Eldridge \& Stanway 2012)
and is observed in local starburst galaxies (e.g. Heckman et
al. 1998,  Leitherer et al. 2011). This method provides a way to probe
statistically the metal content of galaxies and has been used
in z$\sim$3 galaxies (e.g. Savaglio et al. 2004, Sommariva et
al. 2012) and more recently by Faisst et al. (2016) for z$\sim$5
galaxies, however, its use remains limited due to the faintness of the features.
As a result little is known of the metal content of galaxies at very early epochs and
therefore our understanding of the formation of these galaxies
is incomplete.

An alternative way to derive gas phase metallicities relies on the
use of far-infrared (FIR) atomic fine structure (FS) line
transitions, [OIII]52 and 88 $\mu$m, [NIII]57 $\mu$m, [NII]122
and 205 $\mu$m,  which are less susceptible to extinction compared to
optical and UV lines. These lines originate primarily in [HII] regions, their
emission is well understood and can be easily modeled. 
For nearby galaxies these lines are only
accessible from space or stratospheric platforms such as SOFIA (Gehrz
et al., 2009).
The recent advent of the Herschel mission (Pilbratt et
al. 2010) and before that, of ISO (Kessler et al. 1996) provided
measurements of these key FIR FS lines in several nearby galaxies. 
 Herschel was also able to detect FIR FS lines in a handful of
gravitationally lensed sources at redshifts between 2$<$z$<$3.
For distant
galaxies, and, especially those at redshifts z$>$4 where optical line metallicity
diagnostics are not available, the FIR FS lines shift into the
submillimetre range and are now accessible with the Atacama Large
Millimetre Array (ALMA) and the Northern Extended Millimetre Array (NOEMA).

Pereira Santaella et al. 2017 (hereafter PS17) used
the photoionization code CLOUDY (Ferland et al. 2013) to model 
the FIR FS emission lines produced in H\,{\normalsize II} regions as a
function of the metallicity, density, and ionization parameter. The
lines considered were: [OIII] 52 and 88 $\mu$m, [NIII] 57$\mu$m and
[NII] 122 and 205 $\mu$m. A
direct comparison of the observed fine structure lines ratios to model
predictions allow the gas phase metallicity to be determined once a
value for the density and the ionization parameter has been
established. PS17 used FIR FS metallicity ratios to succesfully
determine gas phase metallicities in a sample of local
ultraluminous infrared galaxies (ULIRGs). The
technique outlined in PS17 can be used to estimate the metal content
of heavily dust-obscured galaxies, like the submm-luminous galaxies,
where tranditional methods relying on optical or UV line diagnostics
may fail.


Wide area submillimetre surveys carried out with Herschel, such as the
Herschel Astrophysical Terahertz Large Area Survey (H-ATLAS; Eales et
al. 2010) and the Herschel Multi-Tiered Extragalactic Survey (HerMES;
Oliver et al. 2012), have resulted in the dicovery of large numbers of
dusty star-forming galaxies. HLSW-01 is the brightest gravitationally
lensed source discovered by HerMES. With a magnification of
$\mu$=10.9$\pm$0.7 (Gavazzi et al. 2011), HLSW-01 has been the subject
of significant follow-up effort. SMA interferometric observations
(Conley et al. 2011) resolved the source in four components while  Keck-II$/$NIRC2
K-band imaging resolved the structures further. The NIRC2 data showed
that  the galaxy is magnified by a small group of z$\sim$0.6
galaxies. Based on IFU observations of the brightest foreground galaxy with 
Oxford-SWIFT (Thatte et al. 2006) we measured a spectroscopic redshift
of 0.64 for the brightest member of the group (Foster et al. 2017, in
prep). The redshift of HLSW-01 was established from CO
emission lines as z = 2.958 (Scott et al. 2011, Riechers et
al. 2011). A detailed analysis of the dynamics of the gas showed that
the system is consistent with a gas rich merger at z$\sim$3. More
recent spectroscopic observations with the GTC have confirmed the
merging nature of HLSW-01 (Marques-Chaves et al. 2017, in prep).

In this paper we examine the applicability of the [OIII]88
$\mu$m$/$[NII]122 $\mu$m
line ratio as a metallicity indicator in galaxies near and far. 
We use the aforementioned line ratio to calibrate and derive a
metallicity diagnostic for local normal and star-forming galaxies and
then extend the method to distant galaxies. 
Using
previously unpublished Herschel measurments of FIR FS lines from
HLSW-01 and published data from distant dusty submm-luminous galaxies we
determine, for the first time, their metallicity and examine their
location on the MZR plane. The paper is organised as follows: new
Herschel measurements of the FIR FS lines in HLSW-01 are
presented in Section 2. In Section 3 we summarise the CLOUDY models
and present a new method to estimate the ionization parameter,
the gas electron density
and, determine gas phase metallicities. In Section 4 we apply our method to a
sample of nearby galaxies and to a number of distant submm-luminous
galaxies. In Section 5 we summarize the main findings of this work. 
Throughout this paper, we adopt a cosmology with 
H$_{0}$ = 70 km s$^{-1}$ Mpc$^{-1}$, $\Omega _\Lambda = 0.7$, 
and $\Omega_{m}$ = 0.3.




\section{SPIRE-FTS Observations of HLSW-01}

Observations of HLSW-01 were taken on Ferbuary 20th 2013,  as part
of our Herschel (Pilbratt et al. 2010) OT2 program on far-infrared
spectroscopic observations of 
ultra-luminous infrared galaxies (Rigopoulou et al. 2014). HLSW-01
was observed for 4 hrs with a single
pointing of the SPIRE Fourier Transform Spectrometer (FTS, Griffin et
al. 2010)  sampling across a field of view of 2.6$^{\prime}$ in diameter.
The instrument was used in the high spectral resolution
mode. The
SPIRE-FTS employs two detector arrays covering the 194--313 $\mu$m (SSW)  and 
 303--671 $\mu$m (SLW) wavelength bands simultaneously to measure the 
Fourier transform of the spectrum of a source.
The FWHM
beamwidths of the SSW and SLW arrays vary from 17'' at 194 $\mu$m to 42'' at
671 $\mu$m in a complex fashion due to the nature of the SPIRE
detectors (Swinyard et al. 2014).
\begin{figure*}
	\includegraphics[width=12cm,angle=0]{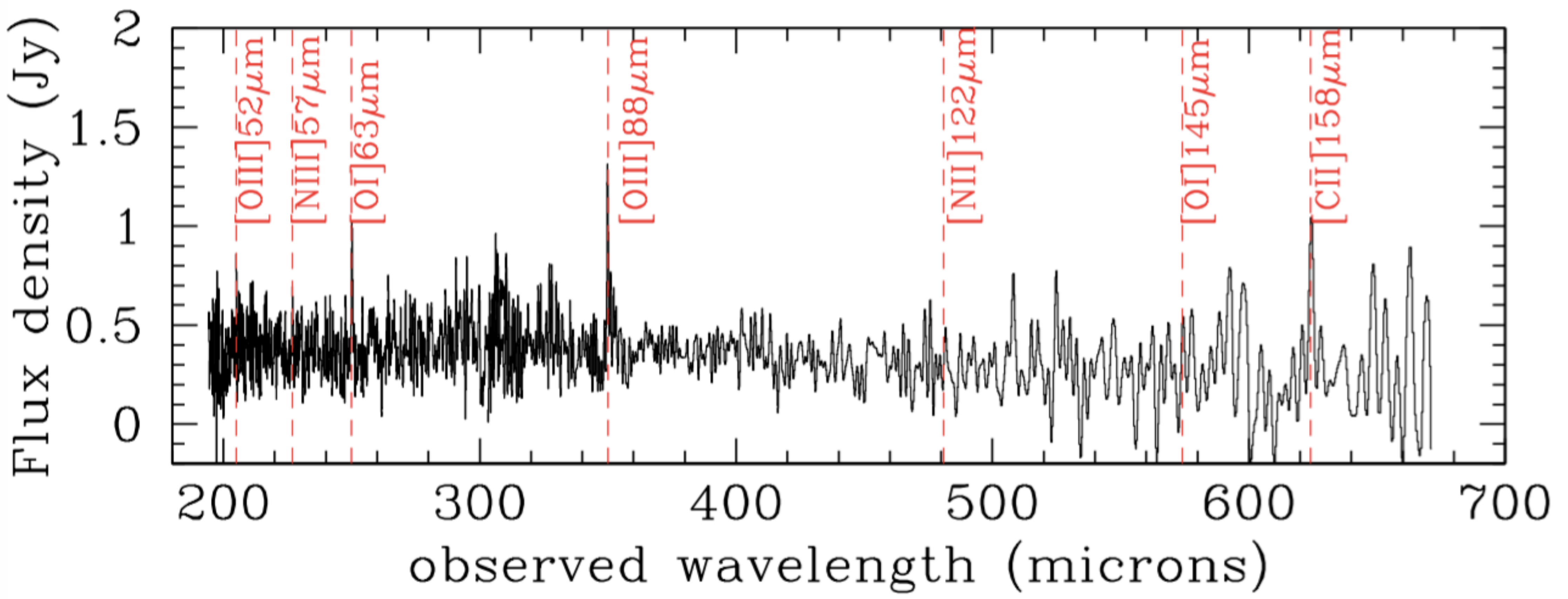}
    \caption{The full SPIRE-FTS spectrum of HLSW-01 in the observed
      frame. The vertical lines correspond the wavelengths of
      [OIII]52, [NIII]57, [OI]63, [OIII]88, [NII]122, [OI]145 and
      [CII]158 microns.  }
    \label{fig:example_figure}
\end{figure*}
 The data were
processed with the SPIRE pipeline (Foulton et al. 2014) in HIPE (Ott et
al. 2010) version 12.1. Since the majority of the targets in our
sample are considered `faint' targets for the FTS post pipeline
processing beyond the standard reduction has been necessary. A
detailed account of the steps followed to reduce the spectra can be
found in Magdis et al. (2014). 
\begin{figure*}
	\includegraphics[width=12cm,angle=0]{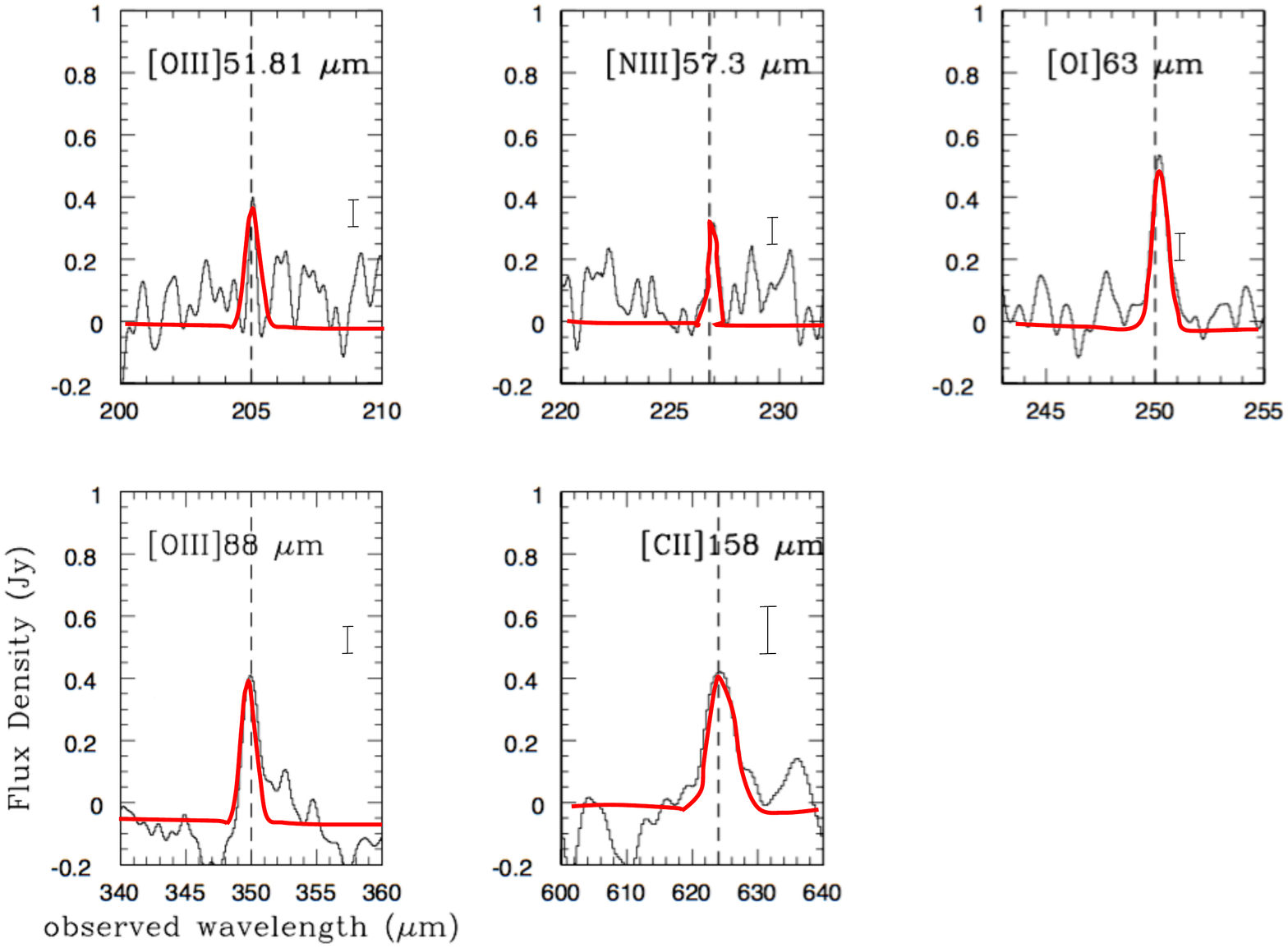}
    \caption{Continuum-subtracted SPIRE-FTS spectra of HLSW-01 around the
      expected wavelengths for (clockwise from top) [OIII]51.81
      $\mu$m, [NIII]57.3 $\mu$m, [OI]63.2 $\mu$m, [OIII]88 $\mu$m and
      [CII]157.7 $\mu$m.The dashed line indicated the expected
      wavelength of each line.  }
    \label{fig:example_figure}
\end{figure*}
We examined the reduced spectra for spectral features at the
positions of expected lines.  To increase the reliability of the line
detections and$/$or minimise the chance of spurious detections we
applied the
Jackknife tecnhique (described in detail in Magdis et al. 2014). Using
this method, decreasing
number of scans (down to ten) are averaged and examined for reccuring
peaks. A real line shows as a consistent peak at the expected
position in all the subsets examined. To measure reliable line fluxes and
background levels bootstrapping was used.  For each observation 
the unaveraged scans were randomly sampled until we reached the
number in the parent population (of 200). These scans were then averaged
and line measurements taken using the same
basic fitting technique as for a standard average spectrum. Baseline
subtraction was performed to each scan prior to resampling. 
 The mean line flux was obtained through resampling of each
observation and then  Gaussian fitting to the resulting line flux.
The Gaussian width
was set at the associated 1$\sigma$ uncertainties. 
The background level above which spectral features were detected was
established from the distributions obtained from the set of random 
frequency positions.
The instrumental line shape of the FTS is well approximated by a sinc
function (see Swinyard et al.2014 and Hopwood et al. 2014 for more
details). For unresolved lines, a sinc function was
fitted simultaneously with a 3$^{\rm rd}$ order polynomial for the
continuum.  If the line was partially resolved then we employed a sinc
convolved with a Gaussian using the SINC-GAUSS function
within HIPE. 
In total we detect the  [OIII]51.81 $\mu$m, [NIII]57.8
$\mu$m, [OI]63 $\mu$m, [OIII]88 $\mu$m and, [CII] 158 $\mu$m lines at
$>$3$\sigma$ while 
for [NII]122 $\mu$m and 
[OI]145$\mu$m lines we report a 3$\sigma$ upper limit. In Table 1 we provide
the lines 
fluxes measured at the expected frquencies. All lines have been
detected at the expected frequency, corresponding to the redshift of
the source of z=2.9574, except for the [OI] 63 $\mu$m line which was
detected at a frequency corresponding to a redshift of z=2.9513.
In Figure 1 we show the full SPIRE-FTS spectrum  while in Figure 2 
we plot the continuum substracted
emission line profiles together with the best fit models. 

\begin{table}
	\centering
	\caption{Line measurements and upper limits for the FIR FS lines
          detected in HLSW-01}
	\label{tab:example}
	\begin{tabular}{lcccc} 
\hline
Line	ID& Obs. Freq. & Flux &FWHM&SNR \\
   $\mu$m &GHz &$\times$10$^{-18}$ W m$^{-2}$& km s$^{-1}$& \\		
\hline\\
 {[}\ion{O}{III}]51.88&1462.60& 6.75$\pm$0.78&293.06&8.0\\
 {[}\ion{N}{III}]57.3&1329.52& 4.42$\pm$1.23&364.03&3.46\\
 {[}\ion{O}{I}]63.2 &1198.95& 8.43$\pm$0.78 &357.32&5.88\\
 {[}\ion{O}{III}]88 & 857.31 &8.38$\pm$1.04&499.83&8.0\\
 {[}\ion{C}{II}]157.7& 480.44&9.06$\pm$2.79 &896.0&3.72\\
\hline\\
Upper Limits & & & &\\
\hline
 {[}\ion{N}{II}]122.1& 618.45&$<$4.08 &425.0&3.0\\
 {[}\ion{O}{I}]145.5& 520.61&$<$2.098 &896.0&3.0\\
\hline\\
\end{tabular}
\begin{flushleft}

\end{flushleft}
\end{table}

 The FIR FS lines provide useful insight into the properties of the ISM
and the underlying ionizing source in HLSW-01. 
Conley et al. (2011) \& Magdis et al. (2014) reported on the unusually warm
colours of HLSW-01 which would imply the presence of an AGN. The [OI]
63 $\mu$m $/$ [CII] 158 $\mu$m ratio can be used to characterise the
nature of the underlying radiation field (e.g. Abel et
al. 2009). Based on the values reported in Table 1 we calculate a ratio
of 0.93 which is more typical of excitation by a starburst rather than
an AGN. On the other hand, the [OIII]88$\mu$m$/$FIR ratio is
1.56$\times$10$^{-3}$. Brauher et al. (2008) found high values of the 
[OIII]88$\mu$m$/$FIR ratio in galaxies with warm 60 $\mu$m$/$100 $\mu$m
colours, attributing this increase either to a higher density of HII
regions or the presence of an AGN. Hence, while HLSW-01 may contain an
AGN it is unlikely that the AGN is contributing significantly to the
FIR FS lines.

\section{FAR-IR fine structure emission line metallicity diagnostics}

\subsection{CLOUDY models and metallicity sensitive FIR FS line ratios}

In Pereira-Santaella et al. (2017, hereafter PS17) we used the spectral synthesis
photoionization code CLOUDY (Ferland et al. 2013) to model 
the far-IR fine structure emission lines produced in HII
regions. We examined how combinations of these far-IR fine structure
lines can be used to constrain gas phase metallicities.

In brief, CLOUDY computes the chemical and thermal structure of a cloud from the
illuminated face to regions of high column density deep into the
clouds where atoms combine to form molecules. CLOUDY incorprorates
photoelecting heating of the gas by grains, cosmic ray ionization and
heating, in addition to photoionization and photodissociation
processes. Details of the processes can be found in van Hoof et
al. (2004), Abel et al. (2005) and in Ferland et al. (2013). 
In PS17 we considered constant pressure models illuminated by a continuous burst of star 
formation. The input spectrum has
been calculated using STARBURST99, Leitherer et al. (1999). Gas phase
abundances were matched to those of the incident spectrum. 
For solar metallicity, the values reported in Asplund et al. (2009)
were used while, for the remaining metallicities we assumed that
abundances scale as Z$_{gas}$/Z$_{\odot}$ for all the elements except
for He and N (values for the latter come from Dopita et
al. 2006). Dust is also included in the models following the
prescription of Remy-Ruyer et al. (2014). A detailed description of
the model assumptions can be found in PS17.

The three most important parameters of the CLOUDY models are the  gas
volume density n$\rm_{H}$ (in units of cm$^{-3}$, hereafter gas density),
the ionization parameter U (defined as the
ratio of hydrogen-ionizing photon density to  gas density,
U=$\phi/$(n$\rm_{H}$c)) and, the gas phase metallicities. 
Constraining the model values for U and n$\rm_{H}$
allows us to determine gas phase metallicities through 
comparison of the observed line ratios to model predictions.

Amongst the various combinations of FIR FS line ratios considered
we found that, the [OIII] to [NIII] ratios are the best metallicity tracers.
This is because the O$^{++}$ and N$^{++}$ lines that dominate
the emission
in HII regions have very similar ionization potentials: 35.12 eV
and 29.6 eV for the [OIII] 52 and 88 $\mu$m and [NIII]57 $\mu$m,
respectively. Because the ionization potentials are rather similar,
their ratio is independent of the specific model value of the
ionization parameter although it shows a mild dependency on the
model value of the gas density. However, the density
dependence of the [OII]52 $\mu$m$/$[NIII]57 $\mu$m ratio is opposite to that of
the [OIII]88$\mu$m$/$[NIII]57 $\mu$m ratio because the [OIII]52 $\mu$m
(88 $\mu$m) is enhanced at high (low) densities. As
demonstrated in PS17, the formula (2.2$\times$[OIII]88+[OIII]52)$/$[NII]57
combining all three lines, reduces the scatter produced by the
dependency on the densities. In section 4.2 we use this ratio to
determine the metallicity 
of  HLSW-01 using the Herschel FIR FS lines presented here.

\subsection{The [OIII]88 to [NII]122 line ratio} 

The [OIII]52 and [NIII]57 $\mu$m lines
are not easily accessible from the ground; only for galaxies at z$>$5 do the
lines shift into windows accessible to ALMA. Here, we consider the [OIII]88$/$[NII]122
line ratio as a potential
gas phase metallicity diagnostic ratio particularly suitable for high
redshift galaxies where both lines shift into the wavelength coverage
of ALMA. The [OIII]88 $\mu$m and [NII]122 $\mu$m lines have similar critical densities, around
300-500 cm$^{-3}$, hence, the [OIII]$/$[NII] ratio does not depend strongly on
the model gas density value. However, the ratio has a strong dependency
on the ionization parameter U since the number of ionizing photons
available will stronly influence the relative amounts of O$^{++}$ and N$^{+}$
gas present. Therefore, in order to use this metallicity diagnostic
ratio we need to determine the value of the ionization parameter U.
In what follows we present a method
that allows us to establish the value of U based on available
continuum measurements.

\subsubsection{The ionization parameter}

Combinations of mid and far-infrared transitions have been found to correlate
with the ionization parameter U. For instance, the
[NeII]12.8 $\mu$m$/$[NeIII]15.5 $\mu$m ratio (e.g. Rigby \& Rieke 2004) or the
[SIV]10.5 $\mu$m$/$[NeIII]15.5 $\mu$m ratio (PS17) have been found to provide good
constraints on the value of the ionization parameter.
However, most of these lines are not accessible with current space or
ground-based facilities. The lines will become within the reach of
JWST but only for $z<2$ galaxies. 

The 60-to-100 $\mu$m flux density ratio (commonly referred to as
C(60$/$100) or colour index) has long been established as a tracer of
the dust temperature T$_{\rm dust}$ of a galaxy (e.g. Chanial et
al. 2007). Furtheromore, it has been shown (e.g. Abel et al. 2009, Fischer et al. 2014)
that, C(60$/$100) varies strongly with the ionization parameter U and,
therefore, can be used to constrain its value. As U increases,
C(60$/$100) increases as well, because the total flux of ionizing photons
goes up. An increase in the total ionizing flux will provide
additional heating to the gas and dust and, therefore, increase the dust
temperature.  Since the dust temperature increases 
the dust emits more energy at shorter wavelengths and the flux at
60 $\mu$m will be higher 
relative to the flux at 100 $\mu$m.  

But although the detailed Spectral Energy Distribution (SED) and the
C(60$/$100) are 
usually widely available for local galaxies this is
not the case for many distant galaxies where only measurements of FIR
FS lines may exist with little or no information on the shape of the
SED of the galaxy. In this case, it is instructive to
investigate whether we can use the 88-to-122 $\mu$m contonuum flux density ratio 
(hereafter C(88$/$122)) to determine the
value of the ionization parameter. To assess whether the C(88$/$122) 
is a viable alternative to C(60$/$100) we first compare the dust temperatures
derived from fitting 60 and 100 $\mu$m and, 88 and
122 $\mu$m continuum flux densities in each case with a modified black body (MBB) with a
fixed dust emissivity index of $\beta$=1.5.  We use the sample of
nearby normal and starburst galaxies from Brauher
et al. (2008) where both sets of measurements, 60 and 100 $\mu$m from
IRAS and, 88 and 122 $\mu$m continuum measurements (derived from ISO-LWS
spectra) are available. For our investigation we select only 
galaxies that are unresolved within the ISO-LWS beam and where
measurements of the continuum at 88 and
122 $\mu$m are available. It is important for the present study that
the galaxies considered are
unresolved within the ISO beam so that line emission from the entire
galaxy is considered. This has the obvious disadvantage that we cannot
explore metallicity gradients however, FIR FS lines are usually not subject to
extinction therefore, the lines could be tracing the emission from all parts of
the galaxy irrespective of extinction. An exception are sources like
Arp 220 where the dust optical depth reaches unity around 200 $\mu$m
and therefore, extinction could become an issue (e.g. Riechers et al. 2013).
In total there are 40 such galaxies where continuum measurements at 88 and 122
$\mu$m are available. The galaxies considered here are all local, the highest redshift 
is z=0.0143. The C(60$/$100) for these galaxies ranges
between 0.32--1.37
while the  C(88$/$122) is in the
range of 0.67--1.97. As mentioned above, we assume a modified
blackbody (MBB) and a dust emissivity of $\beta$=1.5 to convert the
measured C(60$/$100) and C(88$/$122) flux ratios to dust
temperatures T$_{d}$, respectively. In Figure 3 we plot the
temperatures of the MBB fitted to the 60 and 100 $\mu$m fluxes (abscissae)
and 88 and 122 $\mu$m fluxes (ordinates). The standard deviation is
2.8 K which is small enough to validate the use of C(88$/$122) as a
proxy for C(60$/$100) and therefore an estimator of the dust
temperature of galaxies.
\begin{figure}
	\includegraphics[width=6.0cm, angle=-90]{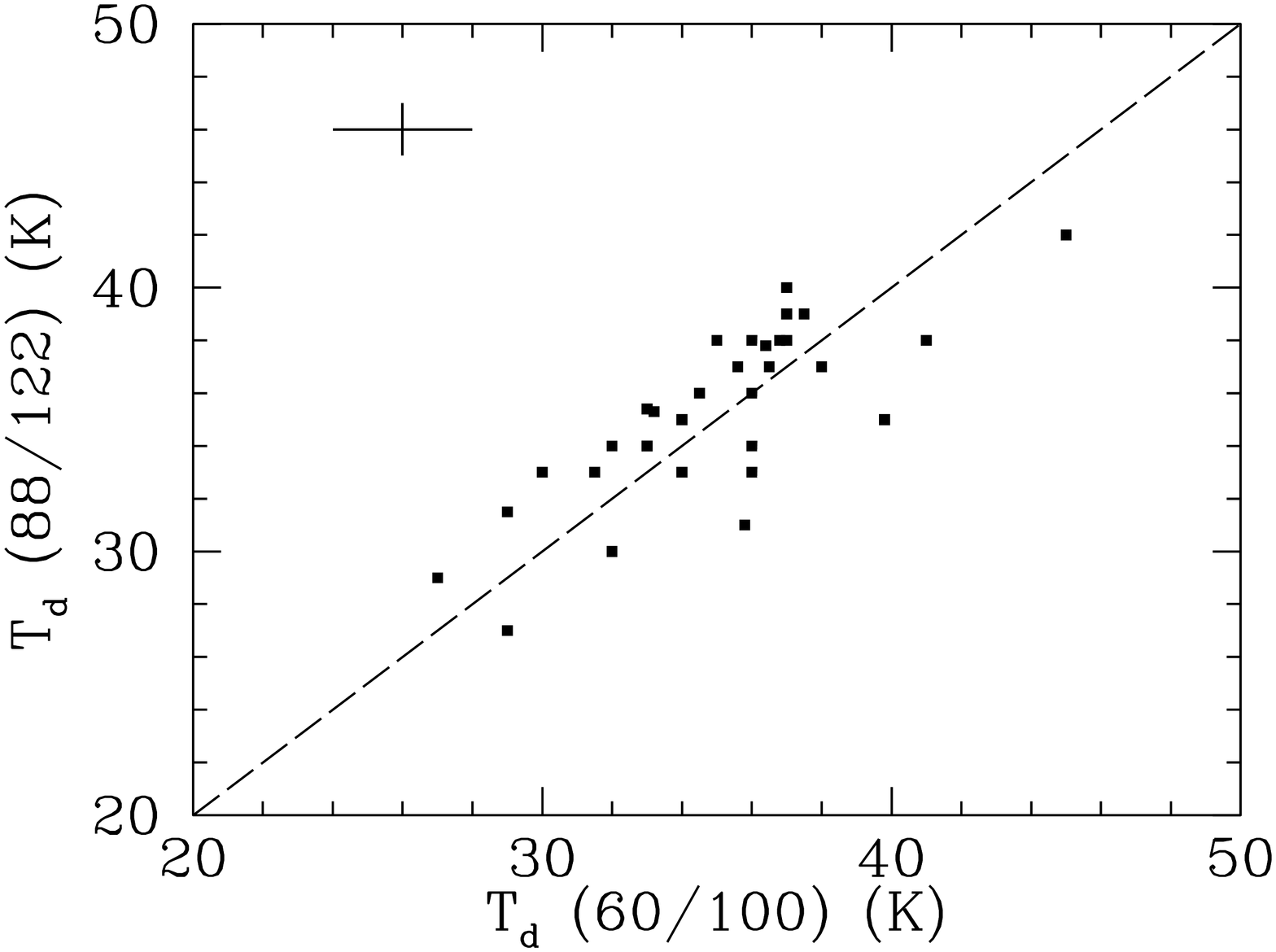}
    \caption{Temperature of the modified blackbodies ($\beta$=1.5)
      fitted to the 60 and 100 $\mu$m flux densities (abscissae) and
      to the 88 and 122 continuum fluxes (ordinates) for the sample of
      local normal and star-forming galaxies with ISO measurements
      (from Brauher et al. 2008). The dashed line
      is the unity line and the average uncertainties are shown in the
    upper left corner.}
    \label{fig:example_figure}
\end{figure}

To further explore how C(88$/$122) varies as a function of the
ionization parameter U, we use the CLOUDY models outlined in section 3.1
to compute the variation of the C(88$/$122) index as a function of the
U parameter. In Figure 4 we plot model values for the C(88$/$122) as a function of
the ionization parameter U for three different values of the  gas density. 
We confirm the trend that the C(88$/$122) increases with increasing
values of U however, we also note that the slope becomes shallower for lower
values of the gas density. The models shown in Figure 4 assume an
A$_{V}\sim$100. This value of A$_{V}$ was chosen since it is
known that extinction toward the most obscured star-forming regions
in galaxies can reach up to such high values (e.g. Abel et al. 2009).
A higher (lower) value of A$_{V}$ will make the
C(88$/$122) ratio grow slower (faster) as a function of increasing
ionization parameter U. 

\begin{figure}
	\includegraphics[width=8.0cm, angle=0]{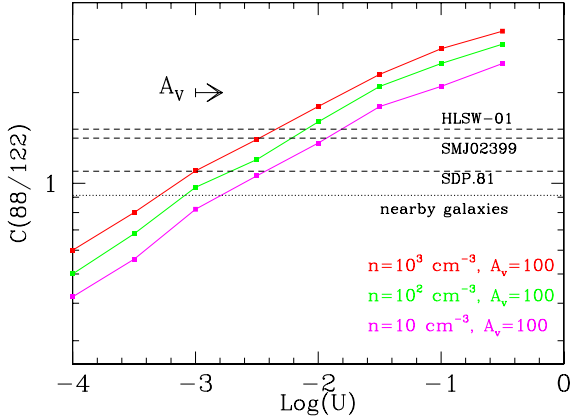}
    \caption{Model predicted  C(88$/$122) flux ratio as a function of the
      logarithm of the ionization parameter, log(U). The plot shows three tracks each for
      different gas density of 10, 100, 1000 cm$^{-3}$ and
      A$_{V}$=100. The dashed lines indicate the  C(88$/$122) for the
    three high-z submm galaxies while the dotted line indicated the
    median value for the ISO sample of nearby galaxies. The effect of
    an increasing A$_{V}$ value is indicated with the arrow.}
    \label{fig:example_figure}
\end{figure}

\subsubsection{The gas density}

The value of the  gas density, n$_{\rm H}$, can
be determined through ratios of mid and$/$or far-IR fine structure
lines from the same ion but with different critical densities. If both
[NII] lines 122 and 205 $\mu$m are available then their ratio provides
a good estimate of the gas density especially at the low density
regime (see Oberst et al. 2006). Alternatively, if no other FIR FS
lines are available a value of $\sim$100
cm$^{-3}$ can be assumed, as has been found by many independent
studies of determining gas densities in other galaxy samples (Section 4.1).

In summary, we have shown that if the
[OIII]88$/$[NII]122 line ratio is used to determine gas phase metallicities
then the ratio of the continuum fluxes C(88$/$122) can be used to
constrain the value of the ionization parameter U. This is
particularly useful for high-z sources where detailed information on
the shape of the SED of the source may not be available.


\section{RESULTS}
\subsection{IR Metallicities of nearby normal and starburst galaxies}

In this section we apply the diagnostics described above to derive
gas phase metallicities of a sample of local normal and star-forming
galaxies and individual high-z submm galaxies with available
measurements of the FIR FS lines. Brauher et
al. (2008) reported measurements of the  [OIII]88 $\mu$m and [NII]122
$\mu$m emission
lines for 30 nearby normal and starburst galaxies (we chose those
galaxies where both lines are detected at S$/$N$>$3).
Following the discussion in Section 3 to determine gas phase metallicities 
through comparison of the observed [OIII]88$/$[NII]122 line ratio to
CLOUDY model predictions, the gas density and the ionization parameter
need to be established. The Brauher et al. paper does
not report measurements of the [NII]205 $\mu$m lines. Instead we
searched the Herschel archive for measurements of the [NII]205 $\mu$m line.
Only 10 of the galaxies with values of the [OIII]88$/$[NII]122 reported in Brauher
et al. have [NII]205 $\mu$m values measured by Herschel. In
deriving the [NII]122$/$205 ratio we applied corrections to account
for the difference in the size of beams (between ISO and Herschel) 
following the procedure
described in Rigopoulou et al. (2013). We measure [NII] 122$/$205
line ratios of 0.6-4  with a median ratio of  $\sim$1.8 corresponding
to an n$_{\rm H}$ value of $\sim$ 60 cm$^{-3}$. 
 A  value of 1 - 300 cm$^{-3}$
with a median value of 30 cm$^{-3}$ was derived by Herrera-Camus et al. (2016) for the
KINGFISH sample of nearby galaxies. PS17 adopted a median value of
n$_{\rm H}\sim $100 cm$^{-3}$, for a sample of ultraluminous infrared galaxies,
derived from the [NII]122$/$[NII]205 $\mu$m
ratio which  is a good probe of the  gas density especially at the low
density limit (e.g. Oberst et al. 2006). We conclude that a value of
n$_{H}\sim$100 cm$^{-3}$ is a good representative value for the gas 
density of normal and starburst galaxies.

Turning to the ionization parameter, in Section 3 we discussed how the continuum
ratio C(88$/$122) can be used to constrain the value of the
ionization parameter. The median C(88$/$122) value for the Brauher
et al.
sample of galaxies is 0.908 (ranging between 0.72 and 1.46) which, for a
 gas density of 100 cm$^{-3}$, corresponds to the logarithm of the ionization
parameter, log(U),  between -3.6 to -2.2. 

Figure 5
shows model predictions for the [OIII]88 $\mu$m$/$[NII]122 $\mu$m line ratio as a
function of metallicity for various values of the ionization parameter
U (the compilation of the model values used can be found in Table A1
of the Appendix of PS17). In the Brauher et al. sample of nearby normal and starburst galaxies the
[OIII]88$/$[NII]122 line ratio varies between 1.15 - 10.63 with a
median value of 2.73. Assuming log(U) values in the
range -3.6$<$log(U)$<$-2.2 the median [OIII]$/$[NII] line ratio of 2.73 implies 
metallicities in the range 0.6$<$Z$_{gas}/Z_{\odot}<$1.0 which
correspond to oxygen abundances of 8.46$<$log(O$/$H)$<$8.69. These 
metallicity estimates are consistent with those based on optical
diagnostics presented for some of these galaxies in Moustakas \&
Kennicutt (2006).

\begin{figure*}
	\includegraphics[width=12.0cm]{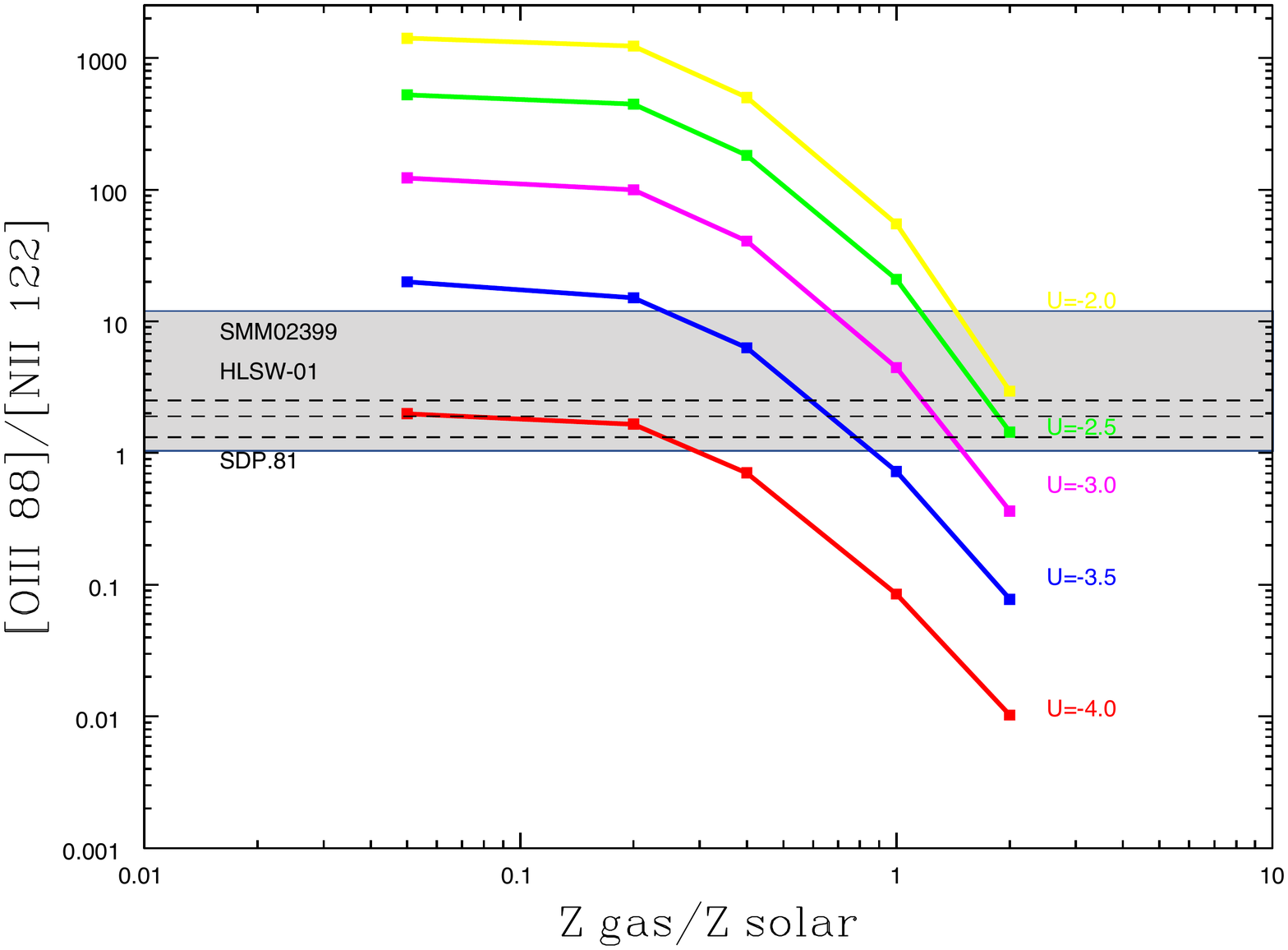}
    \caption{[OIII]88$/$[NII]122 line ratio as a function of
      metallicity. The coloured lines (yellow, green, magenta, blue
      and red) correspond to different values of the ionization
      parameter (log U=-2, -2.5, -3.0, -3.5, -4.0). For each curve we
      consider gas densities of log (n$/$cm$^{-3}$): 1,2,3,4,5.
      The dashed lines correspond
      to the [OIII]88$/$[NII]122 values of SMM02399, HLSW-01 and SPD.81. The
      shaded region corresponds to the values of the Brauher et al. nearby galaxies.}

    \label{fig:example_figure}
\end{figure*}

\subsection{IR metallicities of dusty high-z galaxies} 

Metallicity estimates for large samples of 0.5$<$z$<$2 galaxies have
been derived from R$_{23}$ or from [NII]6584$/$H$_{\alpha}$ ratio
(e.g. Savaglio et al. 2005, Epinat et al. 2009, F\"orster Schreiber et
al. 2009). For galaxies without observations of H$_{\alpha}$ and
H$_{\beta}$ determining the extinction is harder and subject to
assumptions which can influence the metallicity estimates.
Metallicity estimates for z$>3$ galaxies are sparse since the H$_{\alpha}$
and for z$>$4 galaxies the H$_{\beta}$ lines are not observable from
the ground and metallicities have to rely on line ratios involving
[OII]$\lambda$ 3727\.A
and [OIII]$\lambda\lambda$4958\.A and 5007\.A (e.g. Maiolino et al. 2008)  all of which can
be significantly affected by extinction.

The [OIII]88 $\mu$m$/$[NII]122 $\mu$m metallicity diagnostic ratio we
presented above, can provide a suitable
alternative for high-z galaxies, especially for luminous submm galaxies
where line measurements in the rest-frame UV$/$optical regime
are often challening. In what follows we derive metallicities for
three high-z systems where suitable FIR FS lines exist either from
Herschel or ALMA measurements.

\subsubsection{HLSW-01}
Using the FIR FS lines measured by Herschel in HLSW-01
(presented in Section 2) and the methodology outlined above we
determine the gas phase metallicity for this system. 
As we discussed
allready, the two most
critical parameters of the models are the gas density and the
ionization parameter. The [OIII]52$/$[OIII]88
line ratio of 0.805 can be used to estimate the gas density: using Figure 1
(middle) from PS17 we estimate a gas density of 60--100 cm$^{-3}$
corresponding to the [OIII]52$/$[OII]88 line ratio measured for HLSW-01.
The ionization parameter can be inferred from the C(88$/$122) continuum flux
ratio. Using the SPIRE-FTS spectra we measure continuum fluxes
of 318$\pm$10 mJy and 210$\pm$11 mJy around the 88$\mu$m and 122$\mu$m
lines, respectively. We find a value of 1.51 for the C(88$/$122)
ratio. Conley et al. (2011) and Wardlow et al. (2013) have reported
SPIRE continuum measurements for HLSW-01. At the redshift of HLSW-01 of 2.975, 350 $\mu$m
corresponds to rest-frame 88.05 $\mu$m. Likewise, 500 $\mu$m
corresponds to rest-frame 125.8 $\mu$m. Therefore, we can use the
350$\mu$m$/$500 $\mu$m ratio of the (de-boosted) fluxes measured by SPIRE as a proxy for C(88$/$122).
We find that the ratio of the SPIRE fluxes is 1.49, in good agreement
with the C(88$/$122) continuum flux ratio we estimated from the
SPIRE-FTS spectra. The C(88$/$122) ratio of 1.51 corresponds to a 
  log(U)=-2.1
(for an assumed gas density of 100 cm$^{-3}$). 
Using the line fluxes reported in Table 1 we estimate a limit 
for the [OIII]88$/$[NII]122 ratio of $>$2.05. This constrains the
Z$_{gas}$  to be $<$ 0.8-1.0 Z$_{\sun}$ (for n$_{\rm H}$ = 100 cm$^{-3}$). 

Since for HLSW-01 we detected  the 
[OIII]52 \& 88 $\mu$m and [NIII]57 $\mu$m emission lines
(reported in Table 1) we can use the formula
(2.2$\times$[OIII]88$\mu$m$+$[OIII]]52$\mu$m)/[NIII]57$\mu$m,
to get an estimate of the metallicity. The two ratios should provide
similar estimates of the metallicity since they are based on the same
set of models. 
The advantage of using the 
ratio involving the three lines is that the scatter due to the dependency of the
lines on the gas density is significantly reduced (see PS17 for a discussion). 
With a value of 5.44 for the 2.2$\times$[OIII]88$\mu$m$+$[OIII]]52$\mu$m)/[NIII]57$\mu$m
ratio, and assuming an
ionization parameter log(U)=-2.1 and gas density of 100 cm$^{-3}$ we 
find that the gas phase metallicity
for this object is 0.6$<Z_{gas}/Z_{\odot}<$1.0 in accordance with the
limit derived from the [OIII[88$/$[NII]122 ratio. This value of the
metallicity correspnds to
8.46$<12+log(O/H)<$8.69.

\subsubsection{SMJ02399}
The second galaxy for which we derive the metallicity is
SMMJ02399, a luminous submm galaxy at z = 2.803 (Smail et al. 2002)
lensed by the foreground cluster A370. Optical imaging and
spectroscopy by Ivison et al. (1998) originally identified two components,
 a compact component (L1) hosting a narrow-line AGN
(e.g. Villar-Martin et al. 1999), and a second,
diffuse component (L2) associated with L$_{\alpha}$ and H$_{\alpha}$
 emission suggesting the presence of a strong starburst. Further
 analysis of NICMOS imaging (Aguirre et al. 2013) revealed a strong
 diffuse starburst component in L1. 

Using the redshift (z) and Early Universe Spectrometer (ZEUS; Stacey
et al. 2007), Ferkinhnoff et al. (2010) measured [OIII] 88$\mu$m
line emission towards SMMJ02399. Based on the optical
[OIII]5007\.A$/$H$_{\alpha}$ ratio, Ferkinhoff et al. (2010) argued
that the [OIII]88 $\mu$m line emission is consistent with originating in [HII]
regions.  The [NII]122 $\mu$m line towards SMMJ02399 has been 
observed using both ZEUS (Ferkinhoff et al. 2011) and 
ALMA (Ferkinhoff et al. 2015). The [NII]122 $\mu$m line detection with
ALMA is lower than the earlier detection with ZEUS and the authors
attribute the discrepancy to the fact that the [NII] line flux is
resolved out due to significantly more extended emission (and longer 
ALMA baselines) than expected. The spatial extent of the [NII]122
$\mu$m emission supports a starburst origin for the line.
For the current estimates of the
metallicity we use the [NII] 122 $\mu$m line flux reported by
Ferkinhoff et al. (2011) as we are interested in an integrated
measurement of the [NII]122$\mu$m line.
 We conclude that although SMMJ02399 contains an active nucleus (component L1)
both the [OIII]88 $\mu$m and [NII]$\mu$m line emission
originate in the starburst component.
 
We measured the flux ratio C(88$/$122) using the continuum
fluxes at 338 and 463 $\mu$m, which at the redshift of SMMJ02399 corresponds to
rest frame 88 $\mu$m and 122 $\mu$m, respectively. The value of the C(88$/$122) 
was found to be 1.4 which corresponds to values of the ionization
parameter -2.6$<$log(U)$<$-2.1  
assuming gas densities between 10 and 1000 cm$^{-3}$ (since for this
object we cannot constrain the value of the gas density).
The measured [OIII]88$/$[NII]122 line flux ratio of 2.18 then corresponds to a
metallicity of 0.7$<$Z$<$1.1 which translates to
8.53$<$log(O$/$H)$<$8.73.

\subsubsection{SDP.81}
The third galaxy for which we derive an estimate of its metallicity is
H-ATLAS J090311.6$+$003906 (SDP.81)  at a redshift of of  3.042. Valtchanov
et al. (2011) reported a 5$\sigma$ detection for the [OIII]88 $\mu$m line
and and upper limit for the [NIII]122$\mu$m line. Using the SPIRE 350
and 500 $\mu$m continuum fluxes as a proxy for C(88$/$122) we find a
value of 1.09 for the ratio. This value corresponds to a -3.0$<$log(U)$<$-2.6
for values of gas density
in the range 10 to 1000
cm$^{-3}$, respectively. The lower limit for the [OIII]88$/$[NII]122
line ratio is 1.45.
Using this line ratio together with the values of U and $\rm n$,  in
Figure 5, we determine an upper limit for 
the metallicity of SDP.81 of  Z$<$2 Z$_{\odot}$ which corresponds
to log(O$/$H)$<$8.99. 

The finding that the gas phase
metallicities of all three z$\sim$3 submm-luminous sources are consistent
with solar values is in agreement with the normal CO line strengths
found in them and other submm luminous galaxies and the apparent absence of an enhanced
$\alpha_{\rm \small CO}$ conversion factor due to metallicity (e.g. Hodge et al.
al. 2013, Aravena et al. 2016). In the next section, we shall use the FIR FS
metallicity estimates to examine the location of the sources on the mass metallicity plane.

\section{Mass metallicity relation with FIR metallicity diagnostics}

So far, studies of the mass-metallicity relation in nearby and distant
galaxies have relied upon estimates of gas phase metallicities using 
strong line metallicity diagnostics from (rest-frame) optical
emission lines. These line ratios have been calibrated against
metallicities either `directly' (through the use of the dependence of
the metallicity 
on the
electron temperature) or `indirectly' (through photoionization
models). However, it is well known that the output of these methods is
often inconsistent
with each other: metallicity estimates for the same galaxy can vary
widely and introduce artificial evolutionary effects. 
Kewley and Ellison (2008) examined these trends and suggested
conversion factors between different metallicity diagnostics.
In this section we investigate the mass metallicity relation using
new metallicity estimates derived from the FIR FS metallicity
diagnostics.

Figure 6 shows the mass-metallicity relation for the Brauher et
al. (2008) sample of nearby
galaxies. The stellar masses for the nearby
galaxies have been taken from the literature or, using the
mass-to-light ratio from Bell et al. (2003). The (O$/$H) abundance has
been calculated from the metallicity values determined from our models
using the value of 4.9$\times$10$^{-4}$ for the solar
O abundance (Asplund 2009) which corresponds to 12$+$log(O$/$H)=8.69$\pm$0.5.
Although the ISO sample of nearby galaxies contains only
about two dozen galaxies, the plot in Figure 6 already shows some
interesting trends: there seems to be a linear correlation between
stellar mass and metallicity up to about 10$^{9.5}$ M$_{\odot}$ after
which the correlation appears to flatten. The flattening of the
correlation at higher masses is also supported by the best fit to the
data. 
A second order polynomial function provides a better fit (R$^{2}$=0.73) to the data in
comparison to a linear (least square) fit (R$^{2}$=0.47).  The second order
polynomial fit to the data is of the form: 
12$+$log(O$/$H) =-0.098(logM$_{*}$)$^{2}+$2.065(logM$_{*}$)-1.989
where M$_{*}$ represents the stellar mass in units of solar masses. 
\begin{figure}
	\includegraphics[width=9.0cm, angle=0]{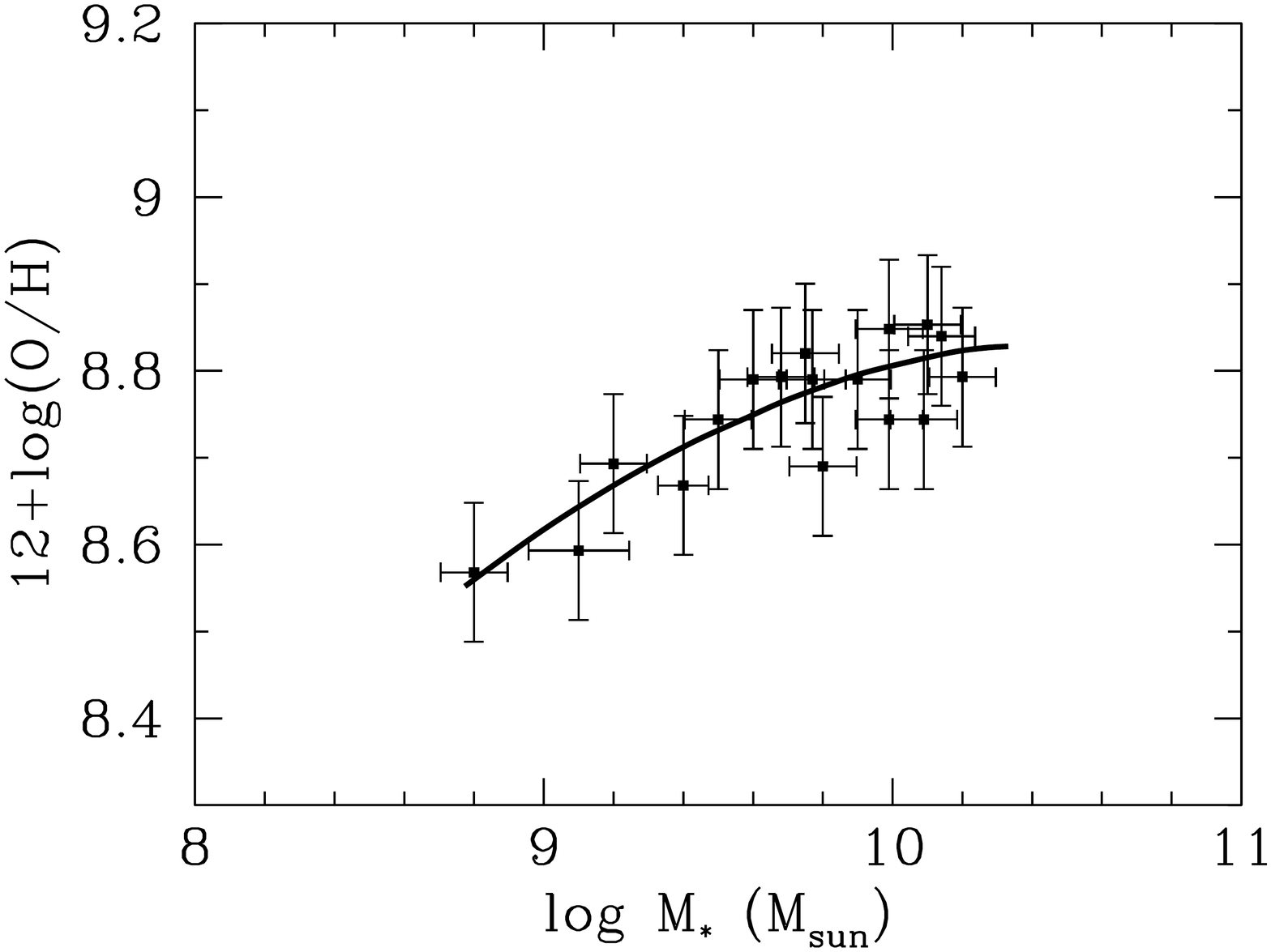}
    \caption{ Mass-metallicity relation for the sample of nearby
      normal and starburst galaxies observed with ISO. The
    best fit mass-metallicity relation derived fro mthe local galaxies
  has the form $y=-0.098x^{2}+2.0654x-1.989$. }
    \label{fig:example_figure}
\end{figure}
Although the ISO galaxies used in this study sample only a limited
range in stellar mass, the fit to the data is in good
agreement with e.g. the Tremonti et al. (2004) calibration both in the
shape of the polynomial as well as the intercept to the y-axis.

\begin{figure*}
	\includegraphics[width=18.0cm, angle=0]{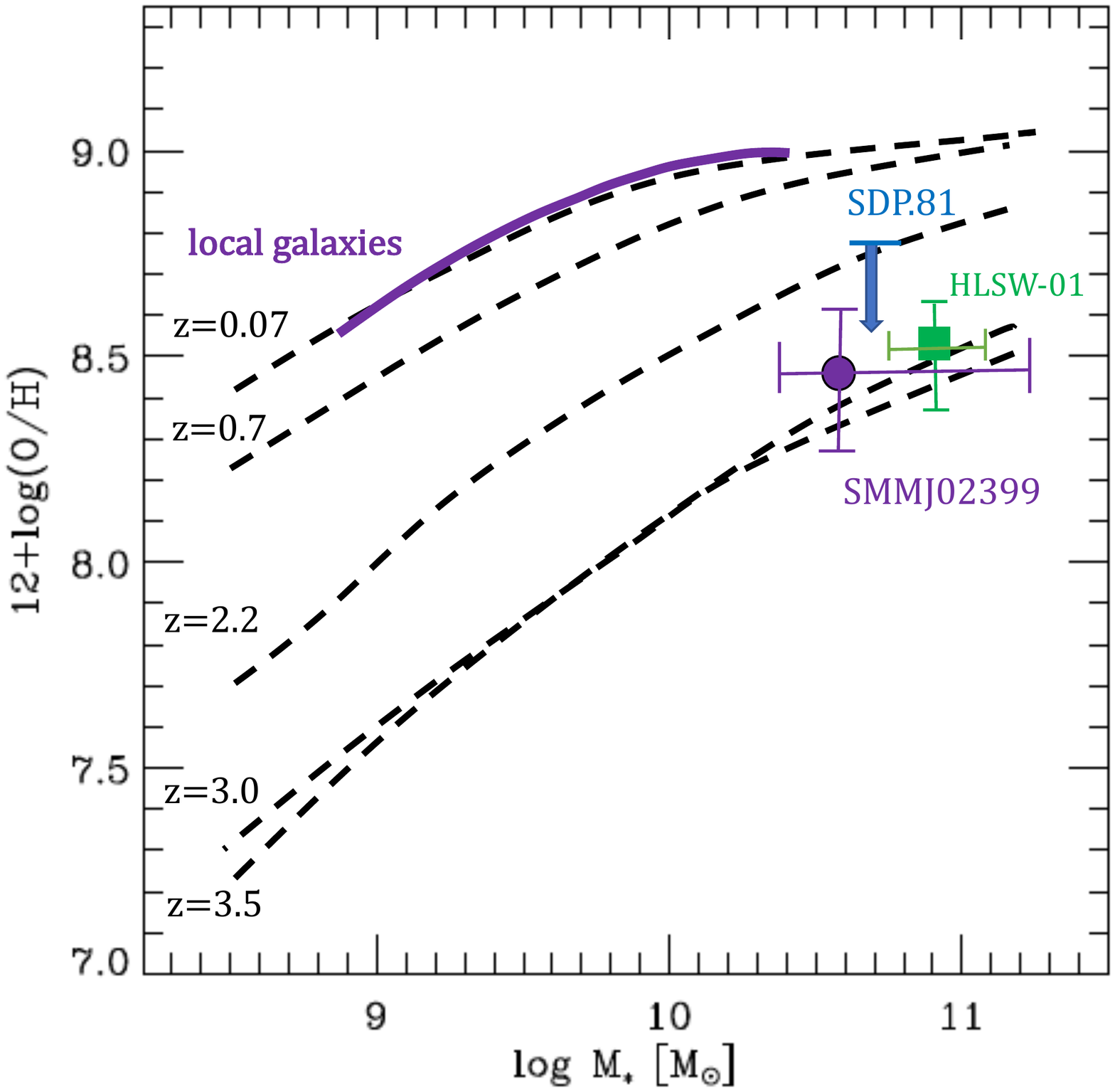}
    \caption{ The mass-metallicity relation for the three high-z submm
    galaxies, HLSW-01, SMMJ03299 and SDP.81. The mass-metallicity
    relation derived in this work based on the local sample of ISO
    galaxies is shown in purple. The z=0.07 relationship comes from
    Kewley \& Ellison (2008) while the z=0.7, 2.2, 3.0 and 3.5
    relationships come from Maiolino et al. 2008 and Mannucci et
    al. 2010. The coloured symbols represent the three submm-galaxies.}
    \label{fig:example_figure}
\end{figure*}

Figure 7 shows the location of the three
z$\sim$3 submm-luminous galaxies for which we estimated their
metallicity based on FIR FS diagnostic ratios on the M$_{*}$-Z plot. 
Stellar masses for the three submm-luminous galaxies have been taken
from the literature: Conley et al. (2011) derived  a stellar mass of
M$_{*}$=(6.3$\pm$3.4)$\times$10$^{10}$($\mu_{L}/$10.9)$^{-1}$
M$_{\odot}$ for HLSW-01. Marques-Chaves et al. (2017, in prep)
reanalysed the SED of HLSW-01 and derived
a revised value for the stellar mass of  M$_{*}$=1.0$\times$10$^{11}$($\mu_{L}/$10.9)$^{-1}$
M$_{\odot}$. Aguirre et al. (2013) reported stellar masses
for the three  SMMJ02399 components: for the MZ plot we use the
stellar mass calculated for the starburst component but the high end
of the error bar
on the stellar mass for this galaxy correspond to the total stellar
mass of the system.
Finally, for SDP.81 Negrello et al. (2014) reported a stellar mass of
M$_{*}$=6.6$\times$10$^{10}$($\mu_{L}/$10.6)$^{-1}$
M$_{\odot}$ where the magnification factor $\mu$=10.6 has been
determined by Dye et al. (2014). 

The vertical error bars show the uncertainty in 12$+$log(O$/$H) from
the model predictions as well as the uncertainty in the measured
[OIII]88$/$[NII]122 line ratio. The horizontal bar shows the
uncertainty in stellar mass (or in the case of SMMJ03299 the range of
stellar mass estimates). The dashed lines in Figure 7 show the
 mass-metallicity relations at
different redshifts taken from the literature, the z=0.07 has been
taken from 
Kewley \& Ellison (2008),
z=0.7, 2.2 and 3.5 from Maiolino et al. (2008) and z=3.0 from Mannucci
et al. (2009). We also show the local mass-metallicity relationship we
derived in Section 4.  The metal content of the  
three z$\sim$3 submm-luminous galaxies to be in accordance with that
expected from the mass-metallicity relation for that particular mass. 

We have shown that FIR FS line ratios provide consistent metallicity
estimates for galaxies near and far. Detection of
[OIII]88 and [NII] 122 $\mu$m emission lines is now possible with ALMA
for galaxies at $z>4$ where optical lines might be out
of reach from the ground.

\section{Conclusions}
Photoionization models for the FIR FS lines found in the spectra
of star-forming galaxies were presented in PS17. Combinations of various FIR FS line
ratios  were
examined to find those that can be used to determine gas phase
metallicities in galaxies. Here, we focused on the applicability of the 
[OIII]88$/${NII]122 line ratio as a potential metallicity diagnostic. 
This diagnostic ratio depends strongly on the value of the 
ionization parameter and only mildly on the value of the gas density. 
We showed that the C(88$/$122) ratio correlates well
with the C(60$/$100) which is sensitive to the ionization parameter
and the shape of the underlying SED. Hence, we suggest that the
C(88$/$122) ratio can be used to constrain the value
of the ionization parameter while the  addition of the [NII]205 $\mu$m emission line
can help constrain the gas density. We applied our methodology to a
sample of local normal and star-forming galaxies and distant z$\sim$3
submm-luminous galaxies. The results of our work can be summarised as follows:

\noindent
(i) The [OIII]88 $\mu$m$/$[NII]122 $\mu$m emission line ratio provides a good
estimate of the gas phase metallicities in a sample of local normal and star-forming
galaxies with ISO FIR FS line measurements. The continuum flux ratio
C(88$/$122) was used to
determine the ionization parameter U while the [NII]
122 $\mu$m$/$205 $\mu$m line ratio was used to estimate gas
densities. We constructed the mass-metallicity relationship and
found that its shape and y-intercept is similar to that derived by e.g.
Tremonti et al. (2004) based on optical emission lines. 

\noindent
(ii) We presented new, previously unpublished, SPIRE-FTS detections of
a number of FIR FS lines from the strongly lensed submm-luminous galaxy HLSW-01 at
z=2.9758. [OIII]52 and 88 $\mu$m, [NIII]57 $\mu$m, [OI]63$\mu$m and
[CII]158 $\mu$m were detected at $>$3$\sigma$ while 3$\sigma$
limits were derived for [OI]145 $\mu$m and [NII]122 $\mu$m. We used the
[OIII]88$/$[NII]122 line ratio and found that the gas phase  metallicity of the
system is close to solar. The solar value for the metallicity of this
system was also found using a second diagnostic ratio combining
[OIII]88, [OIII]52 and
[NIII]57 $\mu$m. 

\noindent
(iii) The [OIII]88$/$[NII]122 line ratio was used to estimate
gas phase metallicities for another two z$\sim$3 submm luminous galaxies
SMMJ0399 and SDP.81 using published data. The metallicities for these
systems are consistent with solar values. We found that all three
z$\sim$3 submm-luminous galaxies studied here exhibit lower metallicities, for a fixed
stellar mass M$_{*}$, compared to the calibration defined for local
(z=0) galaxies. However, the mass-metallicity
relation of the three submm-luminous galaxies agrees well with the
one defined by other samples at the same redshift.

The [OIII]88$/$[NII]122 line ratio can be a potentially powerful
diagnostic of gas phase metallicities for z$>$4 galaxies where most
optical emission lines, traditionally used to establish metallicities,
shift outside of reach from the ground. ALMA can 
observe these FIR FS lines from high redshift systems opening the way
for establishing metallicities in the very early Universe.

\section*{Acknowledgements}

We thank the referee, Graziano Ucci, for his
useful comments and suggestions.
D.R. and M.P-S acknowledge support from STFC
through grant ST/N000919/1. MPS acknowledges support
from the John Fell Oxford University Press (OUP) Research
Fund and the University of Oxford.
IPF acknowledges support from the Spanish Ministerio de Economia y Competitividad (MINECO) 
under grant number ESP2015-65597-C4-4-R.
D.Rie. acknowledges support from the National Science Foundation under
grant number AST-1614213 to Cornell University. SPIRE has been developed by
a consortium of institutes led by Cardiff University (UK) and
including Univ. Lethbridge (Canada); NAOC (China); CEA,
LAM (France); IFSI, Univ. Padua (Italy); IAC (Spain);
Stockholm Observatory (Sweden); Imperial College London,
RAL, UCL-MSSL, UKATC, Univ. Sussex (UK); and Caltech, 
JPL, NHSC, Univ. Colorado (USA). This development has been 
supported by national funding agencies: CSA
(Canada); NAOC (China); CEA, CNES, CNRS (France);
ASI (Italy); MCINN (Spain); SNSB (Sweden); STFC, UKSA
(UK); and NASA (USA). This research has made use of the NASA/IPAC Extragalactic
Database (NED) which is operated by the Jet Propulsion
Laboratory, California Institute of Technology, under contract 
with the National Aeronautics and Space Administration.












\bsp	
\label{lastpage}
\end{document}